\documentclass[nofootinbib,prd]{revtex4}
\usepackage[latin1]{inputenc}
\usepackage{graphicx}
\usepackage{amsmath}

\DeclareFontFamily{OT1}{rsfs}{}
\DeclareFontShape{OT1}{rsfs}{m}{n}{ <-7> rsfs5 <7-10> rsfs7 <10->rsfs10}{}
\DeclareMathAlphabet{\mycal}{OT1}{rsfs}{m}{n}
\def\scri{{\mycal I}}%

\newcommand{\del}{\partial}
\def\eps{\epsilon}
\newcommand{\omegab}{\boldsymbol{\omega}}

\begin{document}
\title{Some aspects of the numerical treatment of the conformal field equations}
\author{Jörg Frauendiener}
\affiliation{Institut für Astronomie und Astrophysik, Universität
  Tübingen, Auf der Morgenstelle 10, 72076 Tübingen, Germany}

\maketitle              

\index{$\scri$|see{scri}}
\index{$\eth$|see{edth}}
\index{initial~data|(}
\section{Introduction}
\label{sec:jf_intro}

The `simulation' of self-gravitating \index{system!isolated}isolated
systems is one of the most challenging and hopefully rewarding tasks
in relativity today. The challenge is to develop a reliable, accurate
and maintainable code. The possible results obtained with this code
may have consequences for different areas of classical relativity such
as the mathematical question of the cosmic censorship hypothesis on
the one hand and the prediction of wave forms for gravitational wave
detectors on the other hand. There exist various approaches towards
the development of such a code (see~\cite{lehner01:_numer} for a
recent review). Their majority is based on (variations of) the
standard ADM equations~\cite{arnowittdeser62}. Some of these numerical
investigations have yielded quite impressive
results~\cite{berger02:_numer_approac_spacet_singul,garfinkle:_simul}.

One of the most successful approaches so far towards the simulation of
isolated systems is based on the characteristic formulation of
Einstein's equation as developed by Bondi~\cite{bondi62:_gravit_vii},
Sachs~\cite{sachs62:_gravit_viii}, Newman and
Penrose~\cite{newman62:_grav_rad_spin_coeff}. This approach is
presented by R.~Bartnik and A.~Norton~\cite{bartniknorton:_numer},
J.A. Font~\cite{font:_local} and L. Lehner~\cite{lehner:_numer_bondi}
in this volume.

The objective of this presentation is to discuss some aspects of the
conformal approach to numerical relativity which is based on Friedrich's
conformal field equations. The general conceptional and analytical
background has been given by Friedrich in this
volume~\cite{friedrich:_confor_einst_evolut} while the numerical
efforts have been summarized in~\cite{frauendiener00:_confor_infin},
see also Husa~\cite{husa:_probl_succes}. Thus, we can take the
opportunity to restrict ourselves to a more detailed discussion of
some of the numerical problems that we encounter when we try to solve
the conformal field equations.

To a large extent the issues involved in the numerical treatment of
the hyperboloidal initial value problem
\index{initial~value~problem!hyperboloidal} for the conformal field
equations are the same as for any system of geometric PDEs which split
into evolution and constraint equations. However, there are some
crucial differences which are due to the fact that null infinity can
be located on the grid which we want to discuss here.

Any numerical implementation of such a system of PDEs will have to
face the following broad issues: $(i)$ the construction of initial
data, $(ii)$ the design of a stable evolution scheme, $(iii)$ the
implementation of (meaningful) boundary conditions, $(iv)$ the control
of the constraint violation during the evolution and, finally, $(v)$
the search for `good' gauges. These subjects are by no means
independent. In fact, the issue $(v)$ is probably the most important
one because it influences all the others. But it is also the most
obscure one because the `goodness' of a gauge can in general not be
inferred without looking at actual evolutions. Hence the `invention'
of good gauges is more like an art than a science.

\index{boundary~conditions|(}
The design of a stable evolution scheme and the design of boundary
conditions are to a large extent uncritical. The choice of boundary
conditions is not as crucial for the conformal approach as it is for
others. This is due to the fact that the boundary of the numerical
grid is located in the unphysical region so that it is causally
disconnected from the physical part of the grid. This means that the
influence of the boundary cannot reach the interior of the
space-time. Clearly, these statements refer only to the analytical
time evolution while the discrete numerical evolution scheme will also
propagate parasitic modes at higher speeds than the speed of light so
that numerically there will be an influence of the boundary on the
physical space-time. However, if we use a well-posed numerical
formulation of the initial-boundary value problem then this influence
should die out with the order of the scheme when we increase the
accuracy.

This implies that we have `only' to make sure that our boundary
conditions are compatible with the numerical evolution scheme in the
sense that the resulting scheme is well-posed. There is a large
literature in the numerical analysis community on the various issues
of boundary conditions (see
e.g.~\cite{engquistmajda77:_absor_bound_cond,%
gustafssonkreiss95:_time,trefethen82:_group}) which can be consulted
for this purpose. The method of choice for the time evolution is a
`method of lines' with fourth-order discretization in space and a high
order Runge-Kutta method to solve the ensuing coupled system of
ODEs. The problem which arises in connection with the boundary
conditions is that of complexity. A high-order numerical evolution
scheme needs boundary conditions discretized at the same order for the
overall scheme to stay at the intended order. This means that one has
to take time derivatives at the boundary and express them using the
evolution equations in terms of the spatial derivatives. This is a
task which can get very complicated for such complex evolution
equations as those obtained from the conformal field equations.

Another issue in relation to the boundary conditions is that they
should be compatible with the constraints. This leads to the necessity
to analyse the the mathematical initial-boundary value problem for the
Einstein and conformal field equations. There is a recent mathematical
result~\cite{friedrichnagy98:_ibvp} which treats the standard Einstein
equations but the extension to the conformal field equations is still
lacking. There is also some related work on the initial-boundary value
problem for linearized
gravity~\cite{szilagygomez00:_cauch,szilagyischmidt02:_bound} and an
implementation of constraint-preserving boundary conditions in
spherical symmetry~\cite{calabreselehner02:_const}.
\index{boundary~conditions|)}

\index{constraints!violation~of|(} The problem of keeping the constraints
satisfied during the numerical evolution is one which seems to plague
most if not all the codes in numerical relativity. The issue here is
that the numerical evolution tends to exponentially drive away the
system from the constraint hypersurface although the analytical time
evolution guarantees that the evolution is confined to the constraint
surface for all times~\cite{gundlachpullin97:_ill}. The question
arises as to whether this exponential violation of the constraints is
a purely numerical effect or whether it has a cause which can be
understood analytically. This cause would have to lie in the way small
deviations from zero in the constraints are propagated by the time
evolution. Analytically this is regulated by the `subsidiary system'
\index{system!subsidiary}of evolution equations for the
constraints. Therefore, in order to shed some light onto this issue
one should analyse the properties of that system in some detail.
\index{constraints!violation~of|)}

There remains the question of finding initial data for the time
evolution. The problems which arise in this area are specific to the
conformal approach and hence we take the opportunity to expose them
here in more detail.

\section{The Andersson--Chru\'sciel--Friedrich procedure}
\label{sec:jf_ach}

\index{conformal!field~equations}
The first step in any numerical simulation is to provide initial data
for the evolution equations. In the conformal approach this means that
we have to determine the fields on the initial hypersurface in such a
way that they obey the conformal constraints which have been obtained
from the conformal field equations by a standard
$3+1$-decomposition~\cite{frauendiener00:_confor_infin}. The structure
of these equations is rather complicated and to this day they have
resisted attempts to cast them into a form suitable for mathematical
analysis or even to devise a reasonable numerical scheme for their
solution. Recently, however, Butscher~\cite{butscher:_explor} has made
some progress towards a reformulation of these equations. Ultimately,
we should strive to obtain initial data directly from solving these
conformal constraints.\index{conformal!constraints}

So far we obtain initial data by an implementation of a procedure
originally due to Andersson, Chru\'sciel and
Friedrich~\cite{anderssonchruscielfriedrich92:_yamab} and expanded on
in~\cite{anderssonchrusciel93:_cauch_einst}.  The mathematical
background and various results are described in detail by
L.~Andersson~\cite{andersson:_conf_constr} in his contribution to the
present volume. We will briefly describe the essential idea. The
starting point for obtaining initial data in \emph{unphysical}
space-time $({M}, {g}_{ab})$ with this procedure is a hyperboloidal
hypersurface $\widetilde\Sigma$ with induced metric $\tilde{h}_{ab}$
and extrinsic curvature $\tilde{k}_{ab}$ in \emph{physical} space-time
$(\tilde{M}, \tilde{g}_{ab})$.

For simplicity we assume that the extrinsic curvature of
$\widetilde\Sigma$ is pure trace, $\tilde{k}_{ab} = \tfrac13
\tilde{h}_{ab} \tilde{k}$, the more general case is treated
in~\cite{anderssonchrusciel93:_cauch_einst}. The vacuum momentum
constraints imply that $\tilde{k}$ is constant and the vacuum
Hamiltonian constraint further implies that the scalar curvature
$\tilde{R}$ is also constant. Rescaling the induced metric by a
constant factor we can assume that
\begin{equation}
\label{eq:jf_R}
\tilde{R}=-6
\end{equation}
which in turn implies $\tilde k = 3$ (recall that a hyperboloidal
hypersurface is characterized by the fact that it has an
asymptotically constant \emph{negative} curvature).

The next step is to assume that the physical space-time admits a
conformal extension so that we may regard $(\tilde{M}, \tilde{g}_{ab})$
as being embedded in the unphysical space-time $({M}, {g}_{ab})$ where
it admits a boundary $\scri$. The two metrics are conformally
equivalent 
\begin{equation}
  \label{eq:jf_metrics}
  g_{ab} = \Omega^2 \tilde{g}_{ab}
\end{equation}
with a conformal factor $\Omega$ which is positive on
$\widetilde\Sigma$ and which vanishes on~$\scri$. The hyperboloidal
hypersurface $\widetilde\Sigma$ extends as a space-like hypersurface
$\Sigma$ beyond~$\scri$ intersecting it transversally.

The embedding of $\Sigma$ in $M$ induces a (negative definite) metric
$h_{ab}$ and extrinsic curvature $k_{ab}$ which are related on
$\widetilde\Sigma$ to the physical data by the following
relations
\begin{equation}
  \label{eq:jf_relat_hk}
  h_{ab} = \Omega^2 \tilde{h}_{ab},\quad \tilde k_{ab} = \frac1\Omega 
  \left( {k}_{ab} - \frac{\sigma}{\Omega} {h}_{ab} \right).
\end{equation}
Here, $\sigma=t^a\nabla_a\Omega$ is the normal derivative of $\Omega$
with respect to the ($g$)-normal vector $t^a$ of~$\Sigma$. This
implies that the trace-free parts transform homogeneously so that if
$\tilde{k}_{ab}$ is pure trace then also $k_{ab}$ is pure trace, while
for the traces we obtain~$\tilde k = k \Omega - 3 \sigma$.

Furthermore, the Ricci tensors of $h_{ab}$ and $\tilde{h}_{ab}$ are
related by the usual conformal transformation formula ($\del_a$ is the
Levi-Civita connection of $h_{ab}$)
\begin{equation}
  \label{eq:jf_rel_curv}
  \tilde{R}_{ab} = R_{ab} - \frac{\del_a \del_b \Omega}{\Omega} -
  h_{ab} \frac{\Delta \Omega}{\Omega} + 2 h_{ab} \frac{\del_c
    \Omega}{\Omega} \frac{\del^c\Omega}{\Omega}
\end{equation}
from which we obtain the transformation of the scalar curvature
\begin{equation}
  \label{eq:jf_relat_sc_curv}
  \tilde{R} = \Omega^2 R - 4 \Omega \Delta \Omega + 6\, \del_c\Omega
  \del^c\Omega .
\end{equation}
Thus, we can express the physical Hamiltonian constraint in terms of
unphysical quantities as
\begin{equation}
  \label{eq:jf_phys_ham}
  \Omega^2 R - 4 \Omega \Delta\Omega + 6\, \del_c\Omega
  \del^c\Omega = -6.
\end{equation}
Recall that the only condition, apart from smoothness, imposed upon
the conformal factor is its vanishing on~$\scri$. Thus, we may write
$\Omega=\omega \phi^\alpha$ for some number $\alpha$ where $\phi$ is a
strictly positive function on $\Sigma$ so that $\omega$ vanishes
on~$\scri$ to the same order as does~$\Omega$. In this way we
`decouple' the two essential properties of $\Omega$: the localization
of $\scri$ on $\Sigma$, indicated by the vanishing of $\omega$ which
we consider to be given once and for all and secondly its role in
relating the physical and unphysical metrics which is taken over by
$\phi$.

Now it is easy to see that upon insertion of this expression for
$\Omega$ into \eqref{eq:jf_phys_ham} one obtains an equation which
contains up to second order derivatives of $\phi$ and, in particular,
terms which are proportional to $\del_c\phi\del^c\phi$. Choosing
$\alpha=-2$ one can eliminate these quadratic terms and arrive at the
equation
\begin{equation}
  \label{eq:jf_yamabe}
  8 \omega^2 \Delta \phi - 8 \omega \del_c \omega \del^c\phi + \left[
    \omega^2 R - 4 \omega \Delta \omega + 6 \del_c \omega \del^c
    \omega \right] \phi = -6 \phi^5.
\end{equation}
Since $\omega$ is fixed we may regard this equation as determining
$\phi$ and hence $\Omega$. This equation could be called the conformal
Lichn\'erowicz equation because it arises from the Hamiltonian
constraint in a similar way as the Lichn\'erowicz equation except that
it is formulated in the unphysical space-time. On the other hand, the
equation determines a conformal factor which fixes a specific member
from a given conformal class of metrics which has constant scalar
curvature. In the mathematical literature this problem is known as the
Yamabe problem. So we call \eqref{eq:jf_yamabe} the
\emph{Lichn\'erowicz--Yamabe--Equation} (LYE). We will discuss this
equation in more detail later on so that we point out only its most
relevant property now: the equation is singular on~$\scri$ in the
sense that for $\omega=0$ all the terms containing derivatives
of~$\phi$ vanish (we assume here that $\phi$ and its derivatives
remain bounded). There remains only an algebraic relation between the
values of~$\phi$ and some geometric quantities. The reason for this
non-regularity is easy to understand. If the equation was regular on
$\scri$ then one would presumably be able to provide boundary data
for~$\phi$. This, however, contradicts the physical picture of~$\scri$
being a universal construction in the sense that for any
asymptotically flat space-time one obtains the `same'~$\scri$. The
degenerate character of the equation on~$\scri$ leads to the question
whether and under what conditions there exist unique solutions. This
has been treated in detail
in~\cite{anderssonchruscielfriedrich92:_yamab} and we will return to
this issue in Sect.~\ref{sec:jf_yamabe}.

In order to find initial data for the conformal field equations we may
thus proceed as follows. We start out with some 3-dimensional manifold
$\Sigma$ carrying a (negative definite) Riemannian metric
$h_{ab}$. Since the equation is conformally invariant it is really
only the conformal class of the metric $h_{ab}$ which is relevant
here. We prescribe the \emph{boundary function} $\omega$ whose sole
purpose is to fix the topology of the space-time we are trying to
simulate. The only condition to be satisfied by this function is that
its zero-set be an embedded 2-dimensional submanifold of $\Sigma$ (it
need not be connected nor have spherical topology). Next we try to
find a non-vanishing solution~$\phi$ of~\eqref{eq:jf_yamabe} with the
given data. This provides a conformal factor $\Omega = \omega
\phi^{-2}$ which conformally rescales the metric $h_{ab}$ to a metric
$\tilde{h}_{ab}$ on $\widetilde\Sigma$ (where $\Omega>0$) with
constant negative curvature $\tilde R = -6$. Imposing $k_{ab} \propto
h_{ab}$ implies $\tilde k _{ab} \propto \tilde h_{ab}$ and so $\tilde
k = const.$ In fact, choosing for $k$ any smooth function on~$\Sigma$
and defining $\sigma=\tfrac13 k \Omega -1$ guarantees that $\tilde k =
3$ so that the physical vacuum constraints will be satisfied
on~$\widetilde\Sigma$.

Once this conformal factor is determined we can continue to generate
the other initial data, such as the Ricci- and Weyl tensors. In the
present case with $k_{ab}$ being pure trace the relevant fields are
the trace-free part of the Ricci tensor and the electric part of the
Weyl tensor. They are given by
\begin{gather}
  \label{eq:jf_ricci}
  \Phi_{ab} = -\frac1\Omega \left( \del_a \del_b \Omega -
    \frac13 h_{ab} \Delta \Omega \right),\\
  \label{eq:jf_weyl}
  E_{ab} = -\frac1\Omega \left(\Phi_{ab} - R_{ab} + \frac13 h_{ab} R\right).
\end{gather}
The crucial point here is that the computation of these tensors
involves \emph{division by}~$\Omega$. This raises two questions: under
what conditions are these expressions well defined and how can we
compute them \emph{accurately} numerically. The first question has
been answered in~\cite{anderssonchruscielfriedrich92:_yamab} while the second
is the subject of Sect.~\ref{sec:jf_initial_data}.

\section{The Lichn\'erowicz--Yamabe--Equation}
\label{sec:jf_yamabe}

\index{Yamabe equation|(}
The degeneracy on~$\scri$ of the
Lichn\'erowicz--Yamabe--Equation~\eqref{eq:jf_yamabe} poses special
problems for its numerical solution. This is due not so much to the
solution process itself which can be handled quite nicely with the
usual methods. The problems arise from the special circumstances
prevailing in the numerical setup. The results
in~\cite{anderssonchruscielfriedrich92:_yamab,andersson:_conf_constr}
apply to the situation where $\Sigma$ is a 3-dimensional manifold
\emph{with boundary given by} $\omega=0$. In this situation there will
be a unique bounded solution of~\eqref{eq:jf_yamabe} on $\Sigma$ which
extends smoothly to the boundary provided the data, i.e. the specified
conformal metric, is smooth up to the boundary and satisfies a
certain condition at the boundary whose physical meaning is the
vanishing of the shear on~$\scri$.

In numerical applications this situation although desirable cannot
usually be achieved. This has several reasons. In 3D simulations one
normally uses Cartesian coordinates while for asymptotically flat
space-times~$\scri$ has spherical `cuts'. This means that the
zero-set of $\omega$ cannot be aligned with the grid and in particular
it cannot be the boundary of the grid. Therefore, there will be nodes
of the grid which lie outside of the physical region,
\emph{beyond}~$\scri$. Even if one decided to adapt the coordinates
to~$\scri$ so that $\omega=0$ lies on a grid plane then this could be
done only for at most two asymptotic regions like in the Kruskal
space-time\footnote{unless one would work with more than one
grid}. For a two black-hole space-time which has three asymptotic
regions one of these would lie inside the grid and there would again
be nodes of the grid which lie beyond~$\scri$ in the unphysical
region. Finally, even in the case of one~$\scri$ and spherical
coordinates one needs for numerical reasons a number of grid points
beyond~$\scri$ in order to implement boundary conditions for the time
evolution. Thus, we have to face the situation where the grid not only
covers the physical space-time but extends beyond~$\scri$ into the
unphysical region.

Therefore, we have to ask what to do at the regions outside the
physical space-time. There are essentially two ways to deal with this
situation. We can either solve the equation only in the physical
region, determine all the initial data there and then extend these
data as smoothly as possible onto the unphysical regions. This is the
path followed by
Hübner~\cite{huebner96:_numer_glob,huebner99:_how_to,huebner99:_scheme_einstein,huebner01:_from_now}.
One major drawback of this possibility is that in general the
conformal constraints will \emph{not be satisfied} in the unphysical
region because the extension does not respect them. Not only is this
an esthetically unpleasing feature -- after all, the unphysical region
can be considered as an asymptotically flat space-time in its own
right -- but almost certainly it will also be counter-productive.
Some features like e.g. radiation extraction and
`scri-freezing'~\cite{frauendiener98:_numer_hivp_i,frauendiener98:_numer_hivp_ii}
\index{scri!freezing} rely on the constraints being satisfied \emph{in
a neighbourhood} of~$\scri$. Furthermore, it is quite likely that the
fact that immediately beyond~$\scri$ the constraints fail could allow
constraint violating modes to penetrate into the physical space-time
triggering the above mentioned exponential growth of the constraints
there (for a discussion of this phenomenon in the context of the
conformal field equations see~\cite{husa:_probl_succes}). Of course,
ideally~$\scri$ should act as a one-way membrane for the propagation
permitting outward flow but inhibiting inward flow. However, this is
true for the (exact) continuous time evolution while for the
(approximate) discrete evolution this depends very much on the method
used to evolve from one time level to the next. It is true that the
influence of the outside region upon the inside should die away with
the order of the evolution scheme employed but ultimately only
numerical tests will settle this point.

So in order to avoid constraint violation \index{constraints!violation~of}
in the unphysical region we
are forced to solve the Lichn\'erowicz--Yamabe--Equation (LYE) on the
entire numerical grid where~$\scri$ is not necessarily aligned with
the grid boundary. However, this is not as straightforward as it may
seem. As alluded to already above, the region outside~$\scri$ can be
regarded as an asymptotically flat space-time in its own right. The
two space-times do not really have any relationship except for the
superficial property that they match on the same numerical grid along
the 2-surface~$\omega=0$. This is reflected in the properties of the
solution of the LYE. For the physical region we know that there is a
unique bounded solution smooth up to~$\scri$ for which we cannot
specify any data. In contrast, in the outside region we have to
specify boundary data \emph{on the grid boundary} in order to obtain a
unique solution. This solution will also be smooth up to~$\omega=0$
but the fact that we can specify arbitrary data suggests that the two
solutions will in general not match smoothly
\emph{across}~$\scri$. The physical reason behind this phenomenon is
simple. We have two asymptotically flat space-times which are forced
to match `at infinity' in such a way that~$\omega=0$ is
the~$\scri^+$ for one space-time and~$\scri^-$ for the other. By
suitably rescaling the metrics we can achieve that the two~$\scri$'s
are isomorphic but there is no reason why the radiation which leaves
one space-time should be the same as the radiation entering the
other. Or, put differently, the mass-aspect obtained from one side is
not necessarily the same as the one obtained from the other side. This
`mismatch' manifests itself in a jump in the third derivative
of~$\phi$. While we cannot influence the part of the solution
inside~$\scri$ we can use the boundary data on the grid boundary to
change the character of the space-time beyond~$\scri$. Thus, with
general boundary data on the grid boundary we obtain a global solution
which agrees with the unique solution inside~$\scri$, which is smooth
for~$\omega\ne0$ but which has a jump in the third derivative
at~$\omega=0$.

This discontinuity will be present also in the initial data computed
from the conformal factor~$\Omega$ and will propagate along~$\scri$
like a shock front moving through a fluid. This is of course not what
we are aiming for. What we need are initial data which are smooth on
the entire grid. To achieve this we need to find a `preferred
extension' beyond~$\scri$ of the unique solution from the inside. This
is determined from the inside by the only requirement of
smoothness. In the case of analytic data this is the unique analytic
extension. The example of the Schwarzschild resp. Kruskal space-time
\index{space-time!Kruskal} \index{space-time!Schwarzschild} given in
the contribution by Schmidt~\cite{schmidt:_data_kruskal} shows that in
general we have to expect that there may be singularities of~$\phi$
beyond~$\scri$ but, hopefully, these will not be immediately adjacent
to null-infinity so that in those cases which are of interest to us
i.e., where the unphysical region is suitably small we can hope that
the function~$\phi$ will be regular on the entire grid.

So what can we do to obtain a sufficiently smooth solution of the LYE
across~$\scri$? Unfortunately, this is not so easy and until now not
solved in a satisfactory way. The first task is to obtain an idea
about the boundary data which are necessary for the smooth
extension. Fortunately, this is straightforward. The degeneracy
of~\eqref{eq:jf_yamabe} yields information about the behaviour of the
solution on~$\scri$. Setting $\omega=0$ in~\eqref{eq:jf_yamabe} gives
\begin{equation}
  \label{eq:jf_phiscri}
  n_an^a \doteq - \phi^4,
\end{equation}
where `$\doteq$' means `equality on $\scri$' and where we have
defined $n_a=\del_a\omega$. Furthermore, assuming that~$\phi$ is
smooth we take a derivative of~\eqref{eq:jf_yamabe} and then put
$\omega=0$. This yields
\begin{equation}
  -2\left(n_e n^e \delta^a_c + \tfrac13 n^an_c \right) \del_a\phi +
  \left( \del_a n_c - \tfrac13 h_{ac} \del^en_e\right) n^a\,\phi \doteq 0.
\end{equation}
Since the map 
\[
x_a \mapsto \left(n_e n^e \delta^a_c + \tfrac13 n^an_c
\right) x_a
\] 
can be inverted this equation and~\eqref{eq:jf_phiscri} allow us to
determine the solution~$\phi$ up to its first derivatives
on~$\scri$. We can use this information to estimate the value
of~$\phi$ on the grid boundary. We arrange~$\omega$ in such a way that
the outer region can be foliated by the leaves~$\omega=const.$ and
introduce a vector field $s^a$ which is transversal to them. This
could be the vector field normal to this foliation but this is not
necessary at this stage. We extend two arbitrary coordinates
$(\xi,\eta)$ defined on~$\scri$ to the outer region up to the boundary
by requiring that they be constant along~$s^a$. In this coordinate
system the grid boundary is given by an equation of the
form~$\omega=\omega(\xi,\eta)$.  Now we have $s^a\del_a \phi= \del
\phi/\del\omega$ and we can estimate the boundary data $\phi_B$ by
\begin{equation}
  \label{eq:jf_bdydata}
  \phi_B(\omega(\xi,\eta), \xi,\eta) \approx \phi(0,\xi,\eta) +
  \frac{\del\phi}{\del\omega} (0,\xi,\eta) \cdot \omega(\xi,\eta).
\end{equation}
Clearly, this estimate can be useful only in those cases where~$\scri$
is close to the grid boundary. Furthermore, it depends very much on
the arbitrary vector field~$s^a$ which is used to set up the
isomorphism between $\scri$ and the grid boundary. There is an
obvious choice for it: the vector field which is normal to~$\scri$
with respect to the freely specified trial metric. Another possibility
is to choose a vector field which is singled out by the numerical
setup and which is transverse to both~$\scri$ and the grid boundary.

With these estimated boundary data we can now proceed to solve the LYE
on the entire grid. Numerically this does not pose a problem. For
demonstration purposes we implemented a simple finite difference
procedure to solve this boundary value problem using
\texttt{Octave}\cite{octav_homep}. Assuming axisymmetry we have the
following ansatz for the free metric
\begin{equation}
  \label{eq:jf_freemetric}
  h_{ab} = -f^2 \left(dx^2 + dy^2 \right) - g^2 r^2 d\phi^2,
\end{equation}
where $f$ and $g$ are arbitrary functions of $(x,y)$ and $r^2 =
x^2+y^2$. It is easy to convince oneself that the general axisymmetric
hypersurface orthogonal metric in three dimensions contains two free
functions. Therefore, the conformal class of these metrics is
characterized by only one free function. Strictly speaking the ansatz
above is too general. We find it rather useful to include this
redundancy because it allows us to test the code on exact
solutions.
\begin{figure}[ht]
\makebox[\hsize]{\hfill\includegraphics[height=50mm]{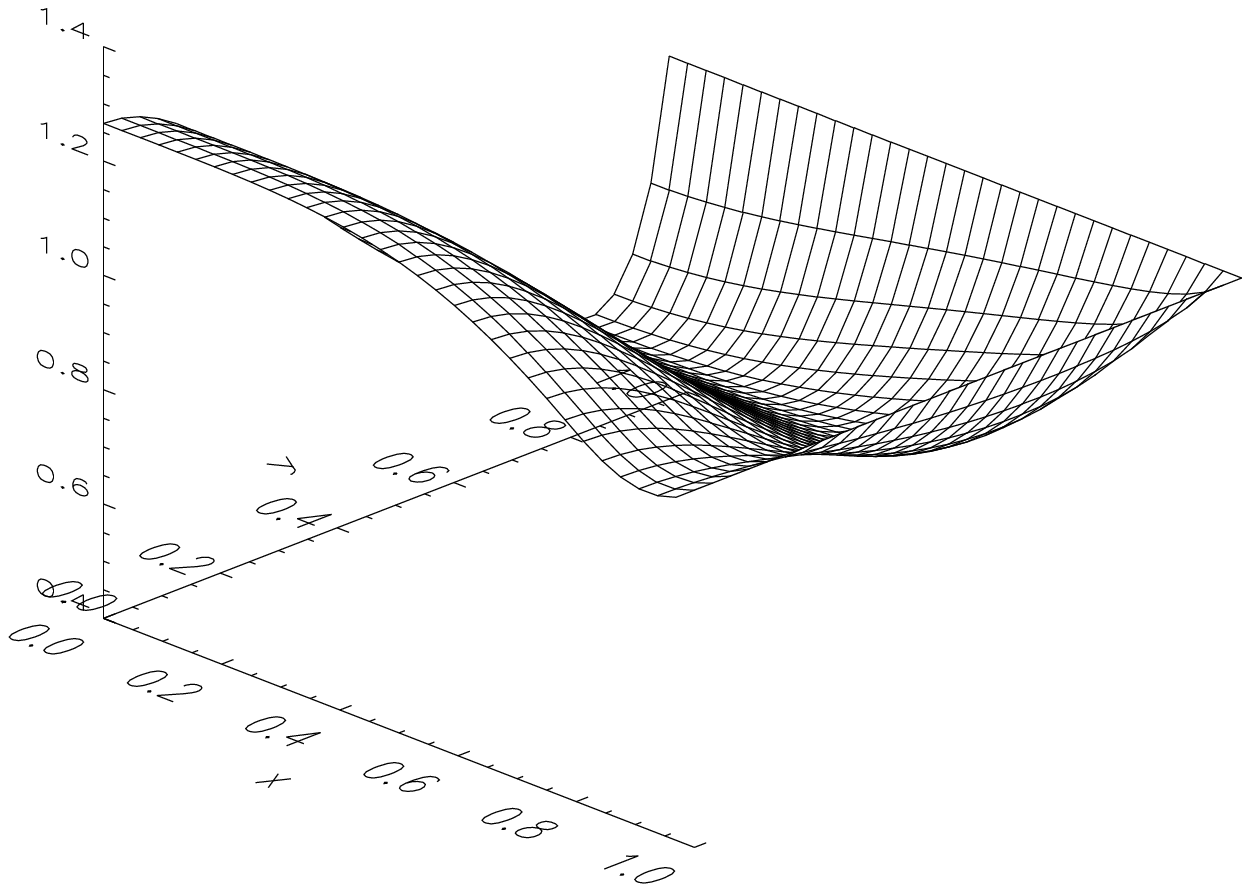}
    \includegraphics[height=50mm]{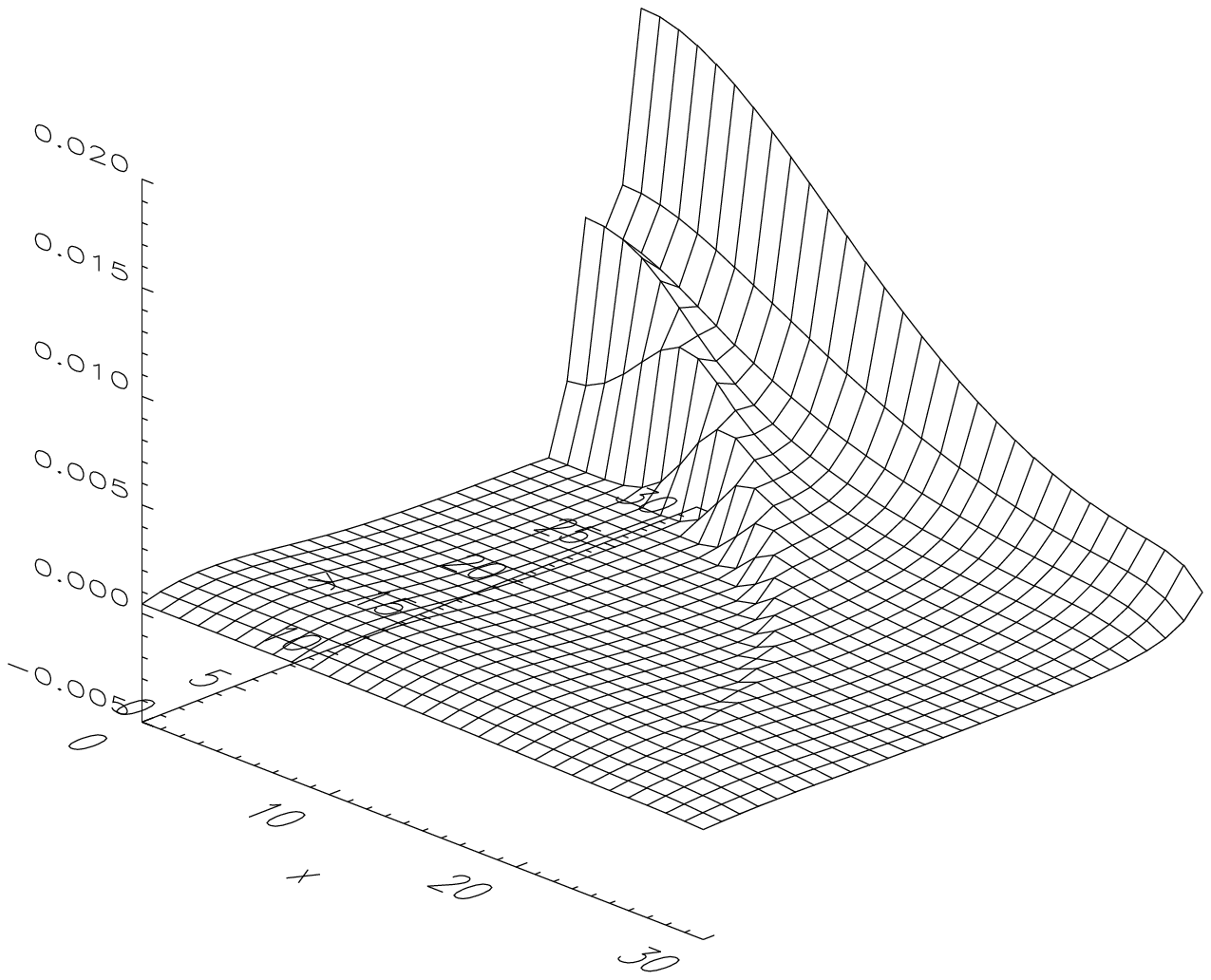}\hfill}
    \caption{\label{fig:jf_u1} Solution (left) of LYE and its third
      $y$ differences (right) for fixed boundary values}
\end{figure}
In~Fig.~\ref{fig:jf_u1} is shown one example of a solution of the LYE
where we have specified $\phi=1$ on the boundary. There is nothing
special about these boundary data except that they are positive. The
solution is obtained on the square $[0,1] \times [0,1]$. The functions
$f$, $g$ are assumed even in both $x$ and $y$. The boundary defining
function is $\omega=r^2-(4/5)^2$ so that $\scri$ is at $r=0.8$. The
diagram on the left shows the solution itself, while the diagram on
the right shows the third differences in the $y$-direction to
illustrate the jump obtained in the third derivative of the solution.

\begin{figure}[ht]
\makebox[\hsize]{\hfill\includegraphics[height=50mm]{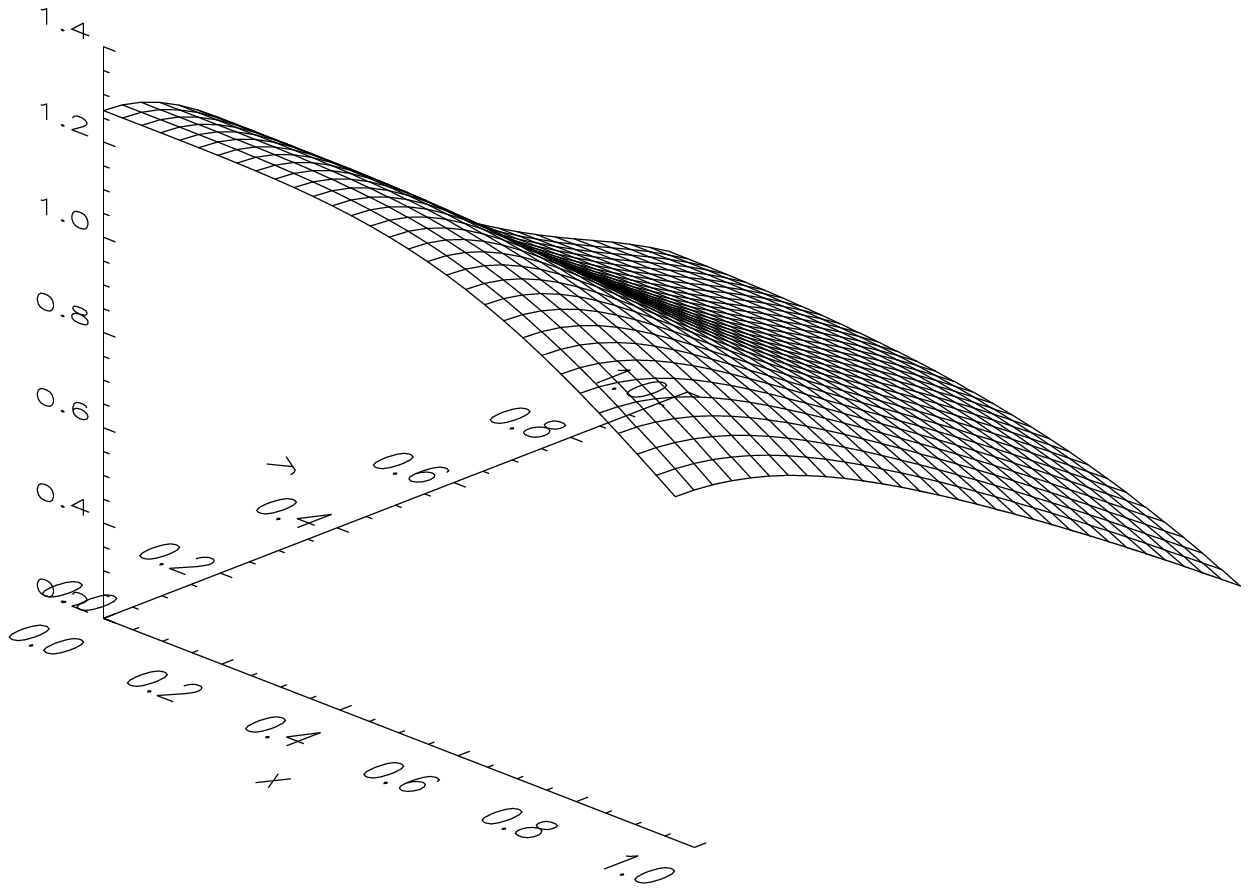}
    \includegraphics[height=50mm]{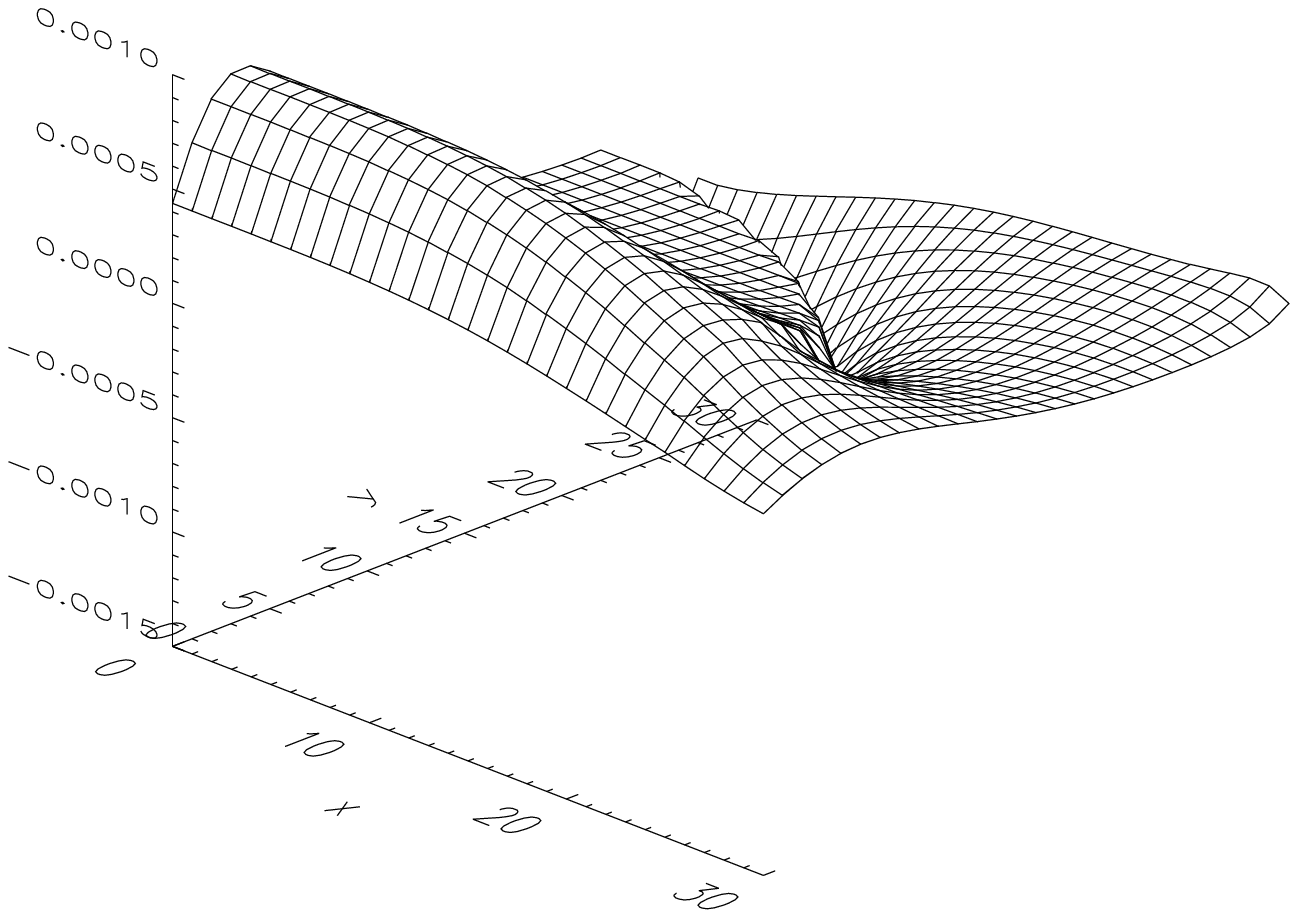}\hfill}
    \caption{\label{fig:jf_u2} Solution (left) of LYE and its third
      $y$ differences (right) for estimated boundary values}
\end{figure}

\begin{figure}[ht]
  \begin{center}
    \includegraphics[height=50mm]{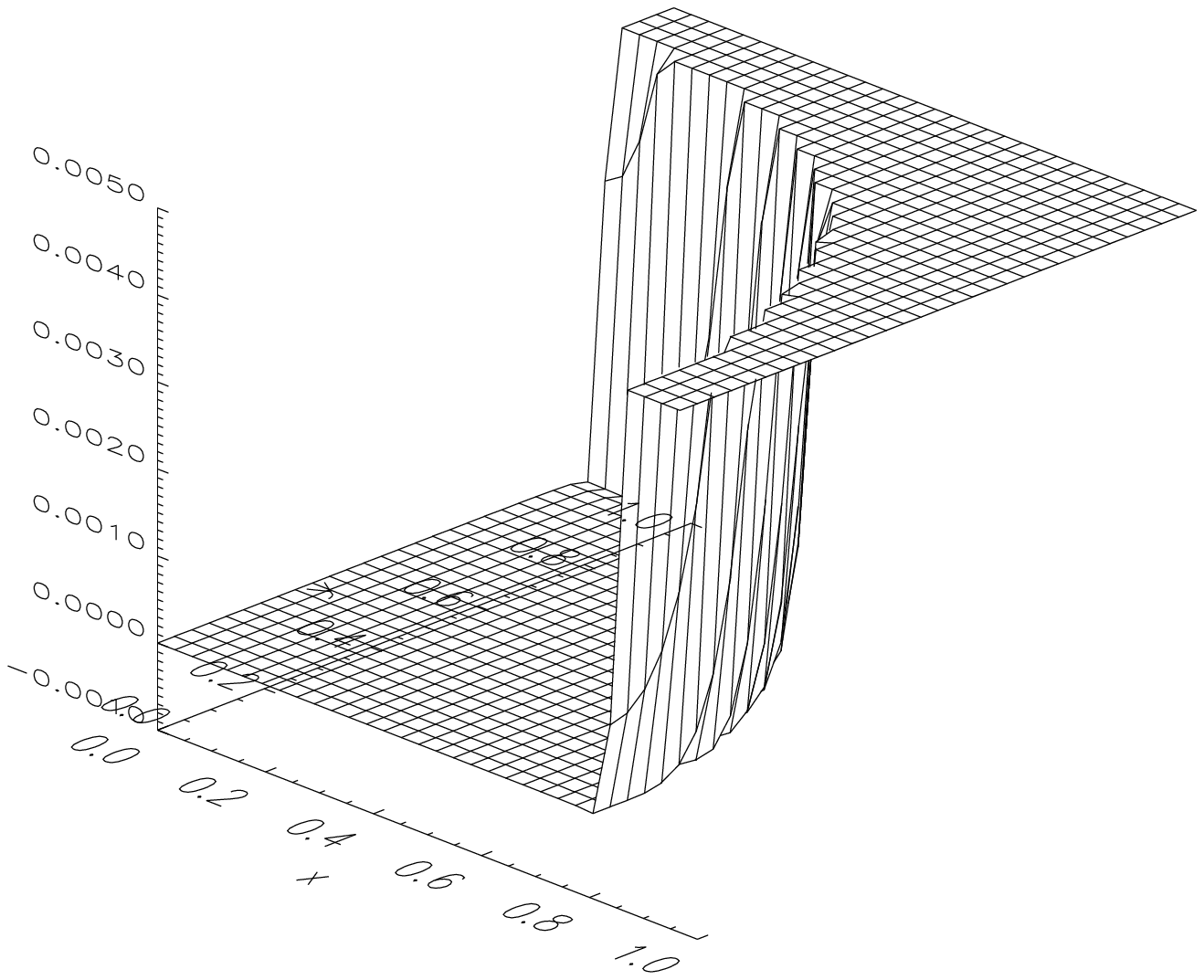}
  \end{center}
  \caption{\label{fig:jf_u12} Difference of the two solutions
    restricted to range $[0,0.005]$}
\end{figure}

In Fig.~\ref{fig:jf_u2}
is shown the solution of the LYE with the same
free data except for the boundary data which have been obtained from
the estimate above. It is clearly visible that the jump in the third
derivative has been greatly reduced (note the different scales in the
two plots) but is still present. As an
illustration we also show in~Fig.~\ref{fig:jf_u12} 
the difference of the two solutions. Here it is clearly visible how
effectively the LYE `shields' the inside from the outside. The
difference of the two solutions in the inside is several orders of
magnitude less than in the outside.

What can be done in order to find a smooth extension of the
solution beyond~$\scri$? One procedure which comes to mind is the
following: We start the solution process by estimating the boundary
data and solve the LYE with these data. Once we are in convergence we
switch from this `inner' iteration to a different, `outer'
iteration. This time we discretize the equation up to the boundary in
the sense that we not only impose the equation at inner points of the
grid but also \emph{on the boundary} by using one-sided
differences. Then we do not need any boundary values. Instead of
prescribing the boundary data in this iteration we fix the value of
the solution at some point in the interior which we know from the
previous iteration. Fixing an interior point is necessary in order to
prohibit the solution to converge to zero which is a valid solution of
the LYE. Once this `outer' iteration has converged we switch back to
the `inner' iteration prescribing the boundary data we have obtained
from the outer iteration. Repeating these steps several times one
hopes to obtain an increasingly smooth solution. This method has not
been fully tested and it remains to be seen to what extent it can
yield useful results.

Another possibility to proceed is to use spectral methods. The reason
for this is the fact that these methods are global on the grid and
there is some hope that this globality helps to obtain a globally
smooth solution. What follows is very much work in progress and so far
there are no results in this direction. However, it is worthwhile to
discuss the possible merits of this approach. 

\index{pseudo-spectral~methods|(}
Spectral methods are described in detail in various
monographs~\cite{boyd01:_cheby_fourier,fornberg96,canutohoussaini88:_spect,gottlieborszag77:_numer_analy,trefethen00:_spect_matlab}
so we can restrict ourselves to the essentials. The general idea is to
use two representations of the grid variables $f$ simultaneously. The
first one is the grid representation where $f$ is given by its values
$f_l = f(\xi_l)$ at some grid points $\xi_l$ while the second one is
obtained by writing $f$ as a finite sum $f(\xi) = \sum_{l=0}^N \hat
f_l P_l(\xi)$ where the functions $P_l(\xi)$ belong to some set of
`basis functions'. The collections $\{f_i\}$, the coordinate
representation, and $\{\hat f_i\}$, the generalized Fourier
representation can both be used to represent the function $f$. The
transformation from one set to the other can be performed by
multiplication with the matrix $M^l{}_m = P_l(\xi_m)$ or its
inverse. The basis functions are chosen according to the specifics of
the problem like symmetry, boundary conditions etc. For some specific
functions like the trigonometric and Chebyshev polynomials the
transformation from one representation to the other can be speeded up
by applying fast transforms like the FFT in various forms.\index{FFT}

In the present problem we look for solutions of the LYE which
are even on $[-1,1]\times[-1,1]$, restricted to the first quadrant
$[0,1]\times[0,1]$.  Thus we may conveniently choose the tensor basis
$\{U_{lm}(x,y):=T_{2l}(x) T_{2m}(y): 0\le l,m \le N\}$ for some
$N$. The functions $T_m$ are the Chebyshev polynomials
\index{Chebyshev~polynomials} of degree
$m$ defined on $[-1,1]$ by the relation
\begin{equation}
  T_m(\cos t) = \cos mt, \qquad \text{for } t\in [0,\pi]. 
\end{equation}
The LYE can be written formally as
\begin{equation}
\label{eq:lyeformal}
  L \phi = - 6\phi^5
\end{equation}
where $L$ is a linear differential operator with coefficients which
vary with the coordinates $(x,y)$. Inserting the ansatz
\begin{equation}
  \phi(x,y) = \sum_{l,m = 0}^N \hat\phi_{lm} U_{lm}(x,y)
\end{equation}
into~\eqref{eq:lyeformal} we find that we need to evaluate the
derivatives of the $U_{lm}$ and multiplicative terms of the form
$f(x,y) U_{lm}(x,y)$ for some functions $f(x,y)$. Here, the advantage
of having both representations at hand is apparent. We can evaluate
the multiplicative terms in the coordinate representation while the
derivatives are efficiently and accurately evaluated in Fourier space
because there exist simple recurrence relations between the
coefficients of a function and those of its derivatives (see any of the
references cited above). Having evaluated these terms and transformed
the result to Fourier space we can regard the linear part of the
Yamabe operator as a linear operator in Fourier space represented by a
$2N \times 2N$-matrix $\hat L$ mapping the coefficients
$\hat\phi_{lm}$ to the coefficients of $L\phi$.

\index{Richardson~iteration|(}
The nonlinear equation is solved by Richardson iteration (see
e.g.~\cite{kelley95:_iterat_method_linear_nonlin_equat}). We compute an
update $\delta \hat\phi$ from a given estimate $\hat\phi_{(n)}$ of the
solution by solving the linear equation
\begin{equation}
  \label{eq:jf_richardson}
  \hat L\, \delta\hat\phi + 30 \hat\phi_{(n)}^4\delta\hat\phi = -
  \left( \hat L\hat\phi_{(n)} + 6 \hat\phi_{(n)}^5\right)
\end{equation}
for $\delta \hat\phi$ and use $\hat\phi_{(n+1)} = \hat\phi_{(n)} +
\delta \hat\phi$ as an improved estimate. However, as it stands the
matrix $\hat L$ is in general degenerate.  When solving the equation
using finite difference methods the same problem appears: the matrix
of the discretized equation is singular. The remedy is, of course, to
add boundary conditions to select a unique element in the kernel. But
it is exactly these boundary conditions which are not known in our
problem and the prescription of which leads to the discontinuity of
the third derivative.

The same applies in Fourier space. In order to get a unique solution
we need to prescribe additional conditions. One aspect of these
additional conditions must be their non-locality in the coordinate
representation i.e., we should not try to fix any particular values of
the solution at grid points. Such conditions would translate into
global conditions in Fourier space. This means that we should impose
\emph{local} conditions in Fourier space e.g. by fixing certain
coefficients. This affects the solution as a whole.

A problem which arises is that in order to prescribe conditions in a
reasonable way we should at least know \emph{how many}. So far this is
not quite clear. It is quite straightforward to convince oneself that
the defect of (the spectral equivalent of) the Laplace operator
obtained in the present setting is $N+1$ and not $2N+1$ as one would
have thought on first sight. The reason is that in the expansion only
\emph{even} polynomials occur and so the operator maps even
polynomials to even polynomials, effectively halving the number of
degrees of freedom. However, $\hat L$ contains not the Laplace
operator but that operator multiplied by $\omega$ together with
various other terms and at least the vanishing of $\omega$ could
enlarge the kernel. Numerical experiments show that this might be the
case but they are inconclusive so far and a rigorous analysis is
needed.

Even if we figure out how many and which conditions to impose we need
to be careful in applying them. For, suppose we solve the linearized
equation subject to some linear condition $B(\delta\hat\phi)=0$ then
the nonlinear solution will necessarily satisfy
$B(\hat\phi)=B(\hat\phi_{(0)})$. Thus, the condition will be
`remembered' in the nonlinear solution and this implies that we have
to impose conditions for the linearized equation without imposing
(restrictive) conditions for the nonlinear equation. Candidates for
such conditions are those which eliminate higher degree polynomials.
The reason is that one expects an exponential decrease in size of the
spectral coefficients of the solution so that these should not
contribute too much.

An entirely different approach is based on a higher order
approximation. Recall that the Richardson iteration scheme is obtained
from the following consideration. Let $N$ be a nonlinear operator
(between finite dimensional spaces for our purposes) and assume we
want to solve the (nonlinear) equation $N[\phi]=0$. In order to derive
the iteration procedure we obtain the equation for an update from an
earlier estimate $\phi$ by putting $N[\phi+\delta\phi]=0$ and
linearizing with respect to the (presumably small) update
$\delta\phi$. We get
\begin{equation}
  \label{eq:jf_N1}
  N[\phi] + dN[\phi] \cdot \delta\phi = 0
\end{equation}
from which the Richardson scheme follows by solving for
$\delta\phi$. Here, $dN[\phi]$ is the linearization of $N$ around
$\phi$. Suppose that $dN[\phi]$ has a non-trivial finite kernel
spanned by $(h_A)_{A=1,\ldots,n}$ for some $n$. Now we have
the problem discussed above that we do not know what to do with the
additional freedom in choosing the homogeneous solutions. However, we
may take the expansion one step further
\begin{equation}
  \label{eq:jf_N2}
N[\phi+\delta\phi] \approx  N[\phi] + dN[\phi] \cdot \delta\phi +
\frac12 d^2N[\phi] \left( \delta\phi,\delta\phi \right).
\end{equation}
Suppose $\delta\phi$ is a solution of~\eqref{eq:jf_N1}. Then so is
$\delta\phi + \sum c_A h_A$. Inserting this
into~\eqref{eq:jf_N2} we obtain
\begin{equation}
  \label{eq:jf_N3}
  \begin{split}
    N[\phi+\delta\phi+\sum_A c_A h_A] = \frac12 d^2N[\phi]&
    \left(\delta\phi,\delta\phi \right) + \sum_A c_A d^2N[\phi]\left(
      \delta\phi, h_A\right) \\
    &+ \frac12 \sum_{A,B} c_A c_B
    d^2N[\phi]\left(h_A,h_B\right).
 \end{split}
\end{equation}
This equation can be considered as a quadratic equation for the
coefficients $c_A$ and we cannot in general expect that this will have
a solution. However, we can expect to find a set $c_A$ so that the
right hand side of~\eqref{eq:jf_N3} is minimal.

The procedure then would be to
\begin{enumerate}
\item take an estimate $\phi$ of the solution
\item compute the linearization $dN[\phi]$ of the nonlinear operator $N$
\item determine a particular solution of~\eqref{eq:jf_N1}
\item determine the kernel of $dN[\phi]$ 
\item compute the second variation $d^2N[\phi]$ and find the set of
  $c_A$ which minimizes~\eqref{eq:jf_N3}
\item finally put $\phi \leftarrow \phi+\sum_A c_A h_A$ and repeat
  from step 1.  until convergence 
\end{enumerate}
Steps~3. and~4. can be performed by finding the \emph{singular value
  decomposition}~\cite{trefethenbau97:_numer_linear_algeb}
\index{singular~value~decomposition} of
$dN[\phi]$. Step~5. is easy in the case where $N$ is the discretized
LYE because then the second variation
of $N$ is simply 
\begin{equation}
(\delta_1\phi,\delta_2\phi) \mapsto 120\phi^3\,(\delta_1\phi)\,(\delta_2\phi).
\end{equation}
This procedure provides a unique update and therefore a unique
solution without imposing any conditions `by hand'. Therefore, one
could hope that this will lead to a sufficiently smooth extension of a
solution of the LYE onto the exterior region. However, numerics is
full of surprises and it is never known whether a theoretically
devised procedure performs satisfactorily in practice.  Work on this
approach has just begun and it remains to be seen to what extent it
is useful.
\index{Yamabe equation|)}
\index{pseudo-spectral~methods|)}
\index{Richardson~iteration|)}

\section{Constructing Initial Data}
\label{sec:jf_initial_data}

\index{division~by~$\Omega$|(}
\index{pseudo-spectral~methods|(}
Now we turn our attention to the next problem which arises in the
construction of initial data for the conformal field equations. As we
discussed above (cf. sect.~\ref{sec:jf_ach}) we have to obtain the
initial data for the Ricci and Weyl tensor once the correct conformal
factor which selects the metric with constant negative curvature from
the given conformal class has been found. The problem we have to face
is the fact that we need to divide two quantities which vanish
somewhere. It is guaranteed from the analytical results that these
quantities vanish simultaneously so that the quotient is well defined
and (at least) continuous. However, numerically at least one of the
quantities is affected by numerical error which implies that it will
not vanish at those places where the other quantity has its zeros.

We write any one of the components in~\eqref{eq:jf_ricci}
or~\eqref{eq:jf_weyl} as $q=C/\omega$ where $C$ contains the specific
component and the factor $\phi^2$ which comes from the conformal
factor $\Omega$ in the denominator. Then we know the denominator (the
freely specified boundary describing function $\omega$) exactly and
its computation at one grid point involves only round-off error. On
the other hand, the numerator $C$ can be computed only up to a certain
accuracy which determined essentially by the truncation error of the
discretization of the LYE. So we have $C = \hat C + \eps$. We denote by
$q=C/\omega$ resp. $\hat q=\hat C/\omega$ the computed and exact
results of the division of the computed resp. exact numerator $C$
resp. $\hat C$.  Let $G=\{x_k\}$ denote the grid and its points on
which we perform the computation. Then we have
\begin{equation}
  \label{eq:jf_estimate}
  |q(x_k) - \hat q(x_k)|= \frac{|\eps(x_k)|}{|\omega(x_k)|}.
\end{equation}
The maximal deviation from the exact result occurs at the grid point
where $|\omega(x_k)|$ attains its minimum. This can be arbitrarily
small, even zero, if $\omega$ happens to vanish at a grid point. If
$|\eps(x_k)|$ would be only due to round-off error as it would be if
we computed the exact analytic solution then this division would not
be disastrous (if we take care to prevent the vanishing of $\omega$ on
any grid point). This is because division is a backward stable process
in IEEE floating point
arithmetic~\cite{stoer72:_einfueh_numer_mathem_i,trefethenbau97:_numer_linear_algeb,higham96:_accur_stabil_numer_algor}.
Then we divide two non-zero numbers which are `simultaneously'
small. The result is a well defined number which approximates the
correct quotient within round-off accuracy.

\begin{figure}[ht]
  \begin{center}
    \makebox[\hsize]{\hfill
      \includegraphics[width=6cm]{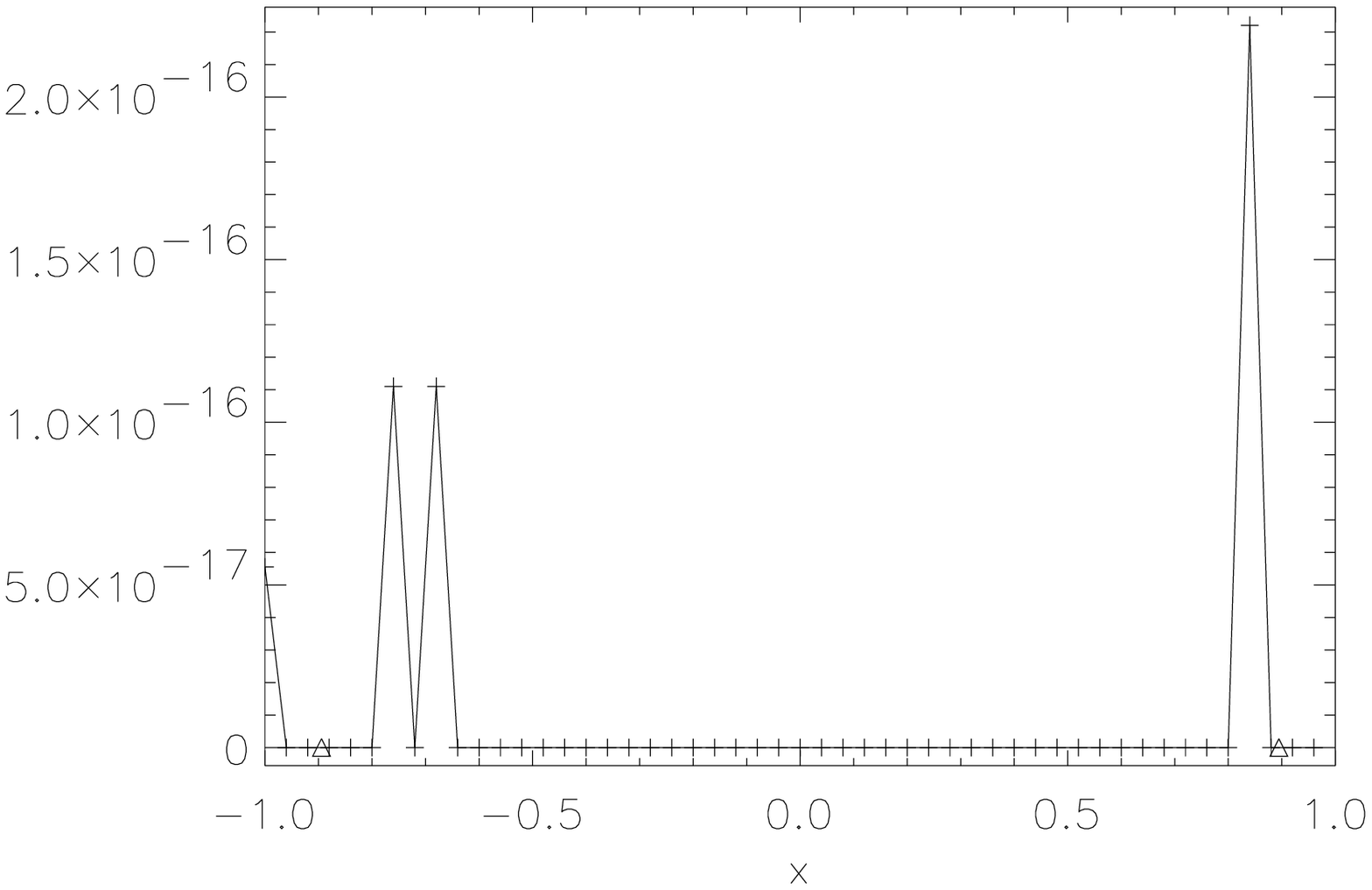}
      \hfill \vline \hfill
      \includegraphics[width=6cm]{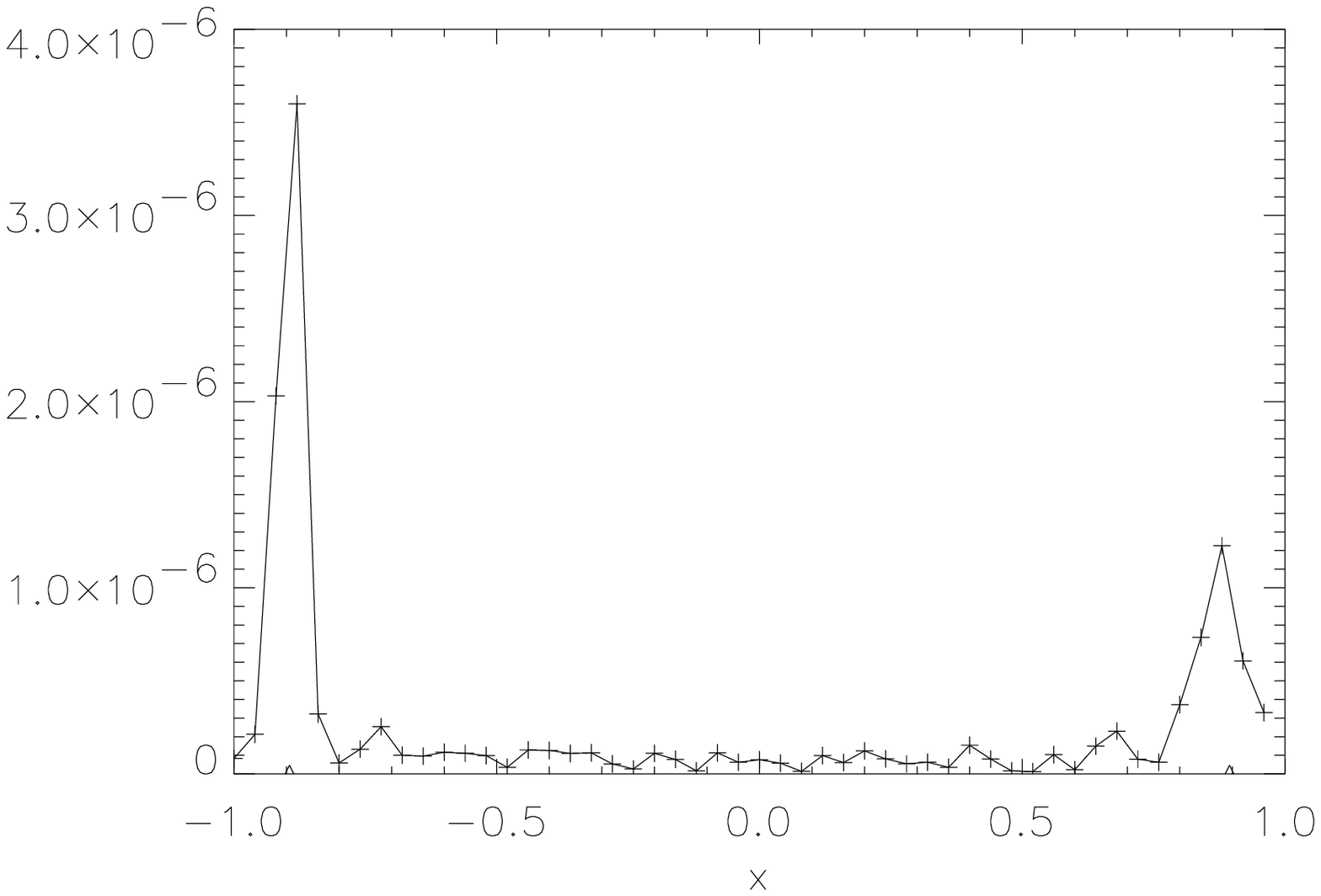}
      \hfill}
    \caption{\label{fig:jf_example} Division by $\omega$ with (right)
      and without (left) additional random perturbation of the numerator}
  \end{center}
\end{figure}

The real disaster occurs when $\eps$ is due to the truncation
error. Then $\eps(x)$ may be orders of magnitude different from the
minimum value of $\omega$ so that the quotient will be wrong by orders
of magnitude. In this situation the numerical implementation of
l'H\^opital's rule to compute the indeterminate term $0/0$ in one form
or another is of no use. The reason is that this rule essentially
replaces the differential quotient by a difference quotient. This
provides a good approximation if one uses the exact expressions for
the numerator. However, if the numerator had been previously computed
with a truncation error then we will also get an answer which is off
by orders of magnitude. To illustrate the problem let us consider a
simple example (cf. Fig.~\ref{fig:jf_example}). Take $\omega = 4/5 -
x^2$ on $[-1,1]$ and let $q$ be some smooth function and define
$C=q\omega$. Thus, $C$ is known exactly and can be computed up to
round-off errors. The diagram on the left shows the error $|C/\omega -
q|$. Clearly, this is on the level of machine precision. The peaks are
not related to the location of the zeros of $\omega$ indicated by the
small triangles on the $x$-axis.

Next we `spoil' $C$ by a random distribution of numbers with maximal
size below $10^{-10}$ to mimic the effect of the truncation error
(although truncation error is certainly not random) and compute again
the error. The result is shown in the diagram on the right. Here, we
see that the overall error is on the level of the random
perturbation. At the location of the zeros the error is maximal.
Thus, we see from this example that \emph{it is the structure of
  $1/\omega$ on the grid  which determines the peaks}.

\begin{figure}[ht]
  \begin{center}
    \setlength{\unitlength}{1mm}
    \begin{picture}(80,60)(-40,-30)
      \thicklines
      \put(-30,-20){\framebox(15,10){\Large $\displaystyle q(x) $}}
      \put( 30,-20){\framebox(15,10){\Large $\displaystyle \tilde q_j $}}
      \put(-30, 20){\framebox(15,10){\Large $\displaystyle C(x) $}}
      \put( 30, 20){\framebox(15,10){\Large $\displaystyle \tilde C_j$}}
      \put(-10, 25){\vector( 1,0){35}}
      \put( 25,-15){\vector(-1,0){35}}
      \put(-22, 18){\vector(0,-1){25}}
      \put( 37, 18){\vector(0,-1){25}}
      \put(  0, 28){\makebox(0,0)[l]{\Large\textbf {FCT}}}
      \put(  0,-18){\makebox(0,0)[l]{\Large\textbf {IFCT}}}
      \put(-20,  8){\makebox(0,0)[l]{\Large $\displaystyle {\cdot \mathbf{\omega}(x)^{-1}}$}}
      \put( 18,  8){\makebox(0,0)[l]{\Large $\displaystyle {\cdot \left(\omega^{-1}\right)^i{}_j }$}}
    \end{picture}
    \caption{The main idea involved in division by $\omega$}
    \label{fig:jf_diag}
  \end{center}
\end{figure}
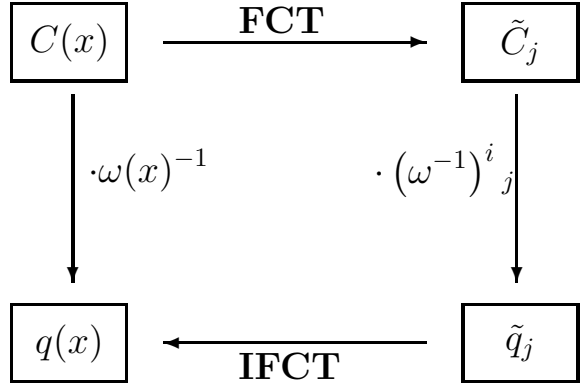

There have been two ways trying to avoid the
disaster. Hübner~\cite{huebner99:_how_to} proposes to solve a
singularly elliptic equation for the quotient in terms of numerator
and denominator. The equation has the same degeneracy on $\scri$ as
the LYE and hence one has to expect the same
problems. In~\cite{frauendiener99:_init_data_ps} we considered a
procedure based on spectral ideas to perform this division. We want to
elaborate on this method here. The main idea, summarized schematically
in Fig.~\ref{fig:jf_diag}
is as follows. We represent the numerator and the denominator both as
a finite sum of basis functions, say Chebyshev polynomials (the
following discussion applies just as well to Fourier series).
Multiplication by $\omega$ is a linear map which we can represent in
Fourier space as a matrix $\omega^i{}_j$ defined by
\begin{equation}
  \label{eq:jf_omega}
  \omega(x) \cdot T_j(x) = \sum_i \omega^i{}_j\,T_i(x).
\end{equation}
Given the representation of $C$ in terms of its coefficients $\tilde
C^j$ then one could simply solve the linear equation $\tilde q^i =
\omega^i{}_j \tilde C^j$ to obtain the coefficients of the
quotient. However, there are some snags. 

The basis functions $\{T_j: 0 \le j \le N\}$ span a
\emph{finite}-dimensional vector space $V_N$. Since multiplication
with any polynomial (except $T_0$) increases the degree we should
obtain products with degree higher than $N$. However, if we regard the
multiplication map as a map from $V_N$ into itself we are asking to
represent a high degree ($>N$) polynomial as a sum of polynomials of
degree up to $N$. With this `folding back' of high degree polynomials
into the range of low degree polynomials -- this effect is known as
\emph{aliasing} -- multiplication is no longer injective. Therefore,
we need to enlarge the target space. Multiplying two polynomials of
degree at most $N$ yields a product of degree at most $2N$. So we
regard multiplication by $\omega$ as a map $\omega: V_N \to
V_{2N}$. This map is injective so that the image $\omega[V_N]$ of
$V_N$ is a $N+1$-dimensional subspace in $V_{2N}$. The matrix
$\omegab$ of the
map $\omega$ is no longer square but has dimensions $(2N+1) \times N$.

Let $\tilde C^j$ be the representation of $C=\sum_{j=0}^N \tilde C^j
T_j$ in $V_N$ which is trivially also a representation in $V_{2N}$. In
general, $\tilde C^j$ will not lie in $\omega[V_N]$ because the error
in its computation will drive it away. Hence, the first step
in the computation of the quotient is to project $\tilde C$ onto
$\omega[V_N]$ and then the second step is the inversion of $\omega$ on
its image.

\index{QR~factorisation|(}
Both these steps can be accomplished by computing the (reduced)
QR-factorization of $\omegab$ 
(cf.~\cite{trefethenbau97:_numer_linear_algeb}). This algorithm allows
us to write the matrix $\omegab$ uniquely as $\omegab = QR$ with a
square $(N+1) \times (N+1)$ upper triangular matrix $R$ and a $(2N+1)
\times (N+1)$ matrix $Q$ whose columns are orthonormal. The crucial
fact for us is that these columns span $\omega[V_N]$. This implies
that $P = QQ^t$ is a projector onto that space while $Q^tQ = id_{N}$
is the identity on $V_N$.

Thus, to solve the equation $\tilde q = \omegab \tilde C$ we proceed
in two steps: first compute $\tilde z = Q^t \tilde C$ and then solve
$\tilde q = R^{-1} \tilde z$. These steps correspond exactly to the
two steps mentioned above because $Q\tilde z = P\tilde C$ is the
projection of $\hat C$ onto $\omega[V_N]$.  Since $R$ is the matrix
representation of $\omegab$ with respect to the basis provided by the
columns of $Q$ and its pre-image in $V_N$ its inverse provides that
vector whose image under $\omegab$ is the projection of $\tilde C$
onto $\omega[V_N]$.

The scalar product involved in the computation of the QR-factorisation
is the natural scalar product between the Chebyshev polynomials. Hence
the projector is the orthogonal projector onto $\omega[V_N]$ with
respect to this scalar product. The projection $P\tilde C$ is that vector
in $\omega[V_N]$ which is closest to $\tilde C$ with respect to that
scalar product. This is not what is needed here. For suppose that
$\omega$ is a polynomial of degree $M<N$. Then the projection should
not contain any polynomials up to degree $M-1$ because any of these
would give rise to a singular quotient. Thus, we need to construct a
different projection $\hat P$ which has the same image as $P$ but
which annihilates all polynomials with degree less than $M$.

In order to do this we have to impose the condition that $\omega$ be a
polynomial with degree $M<N$. This is not a severe restriction because
$\omega$ can be specified freely. Then we consider $\omegab : V_N \to
V_{N+M}$ as a map into $V_{N+M}$. To find the new projection it is
necessary now to compute the full QR-factorisation of $\omegab$
because this provides not only information about the image of
$\omegab$ but also about its orthogonal complement. In fact, one can
again write $\omegab = \hat Q \hat R$ where now $\hat Q$ is an
orthogonal $(N+M+1)\times(N+M+1)$ matrix and $\hat R$ is upper
triangular with dimensions $(N+M+1)\times(N+1)$. Again the first $N+1$
columns of $\hat Q$ form an orthonormal basis of $\omega[V_N]$ while
the remaining columns span its orthogonal complement, the kernel $K$
of $P=QQ^t$, so $K = \ker P$. We denote by $\hat Q_1$ the matrix
containing the first $N+1$ columns of $\hat Q$ and by $\hat Q_2$ the
remaining ones. Then $\hat Q = \bigl(\hat Q_1 | \hat Q_2\bigr)$. Note,
that $\hat Q_1$ coincides with the matrix $Q$ obtained above from the
reduced QR-factorisation.
\index{QR~factorisation|)}

\begin{figure}[ht]
  \begin{center}
        \makebox[\hsize]{\hfill
      \includegraphics[width=5cm]{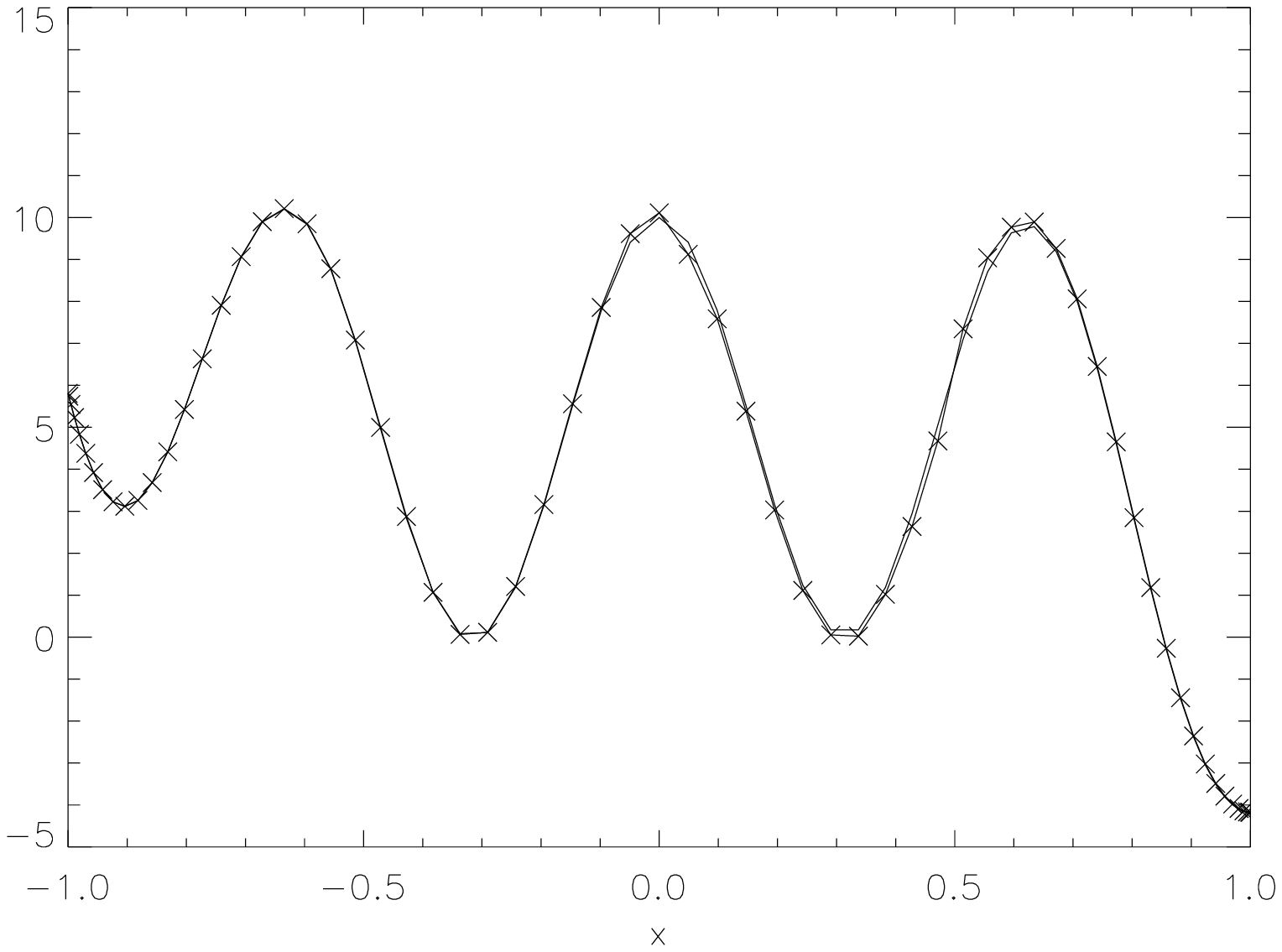}
      \hfill \vline \hfill
      \includegraphics[width=5cm]{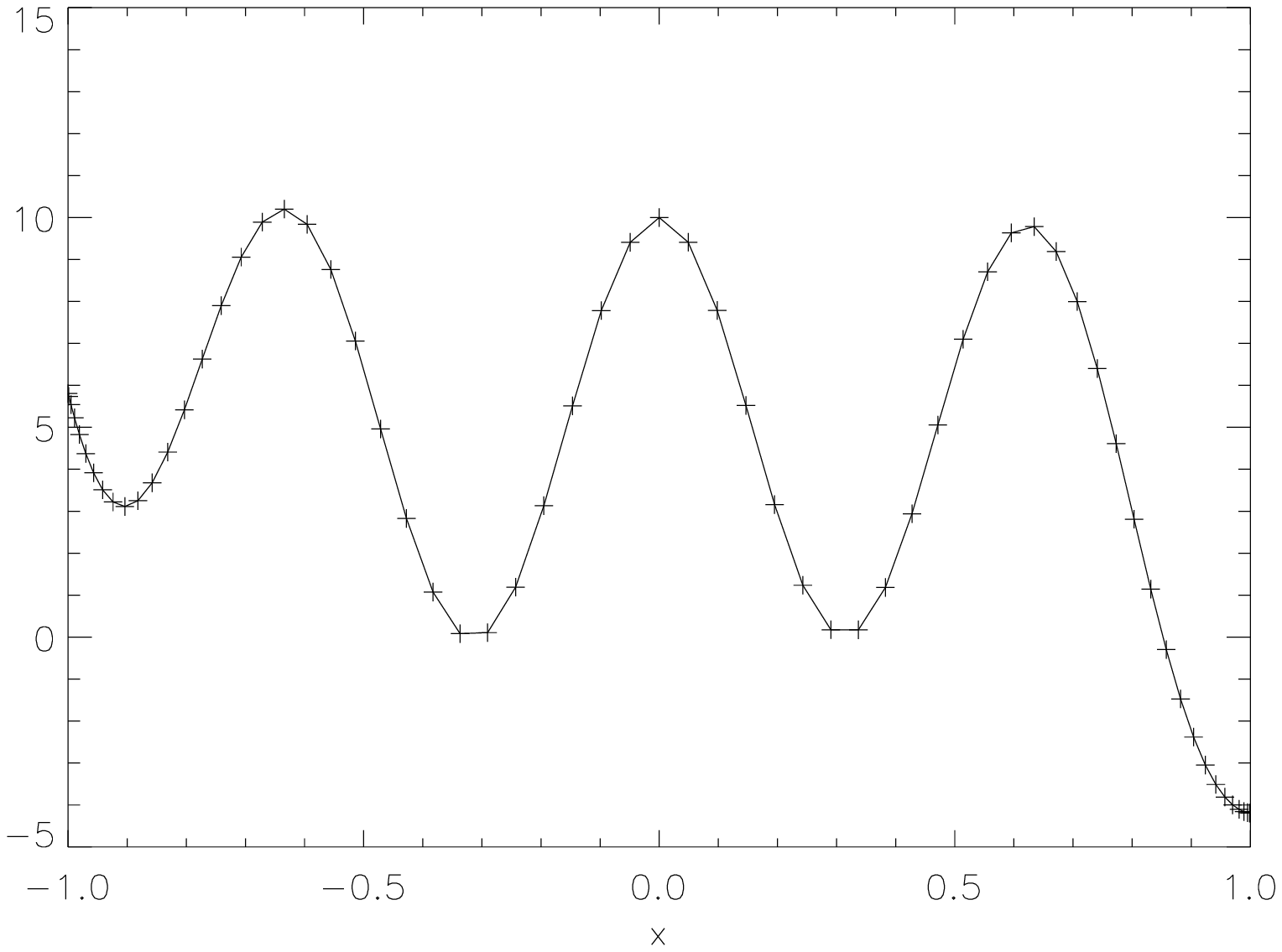}
      \hfill}
    \makebox[\hsize]{\hfill
      \includegraphics[width=5cm]{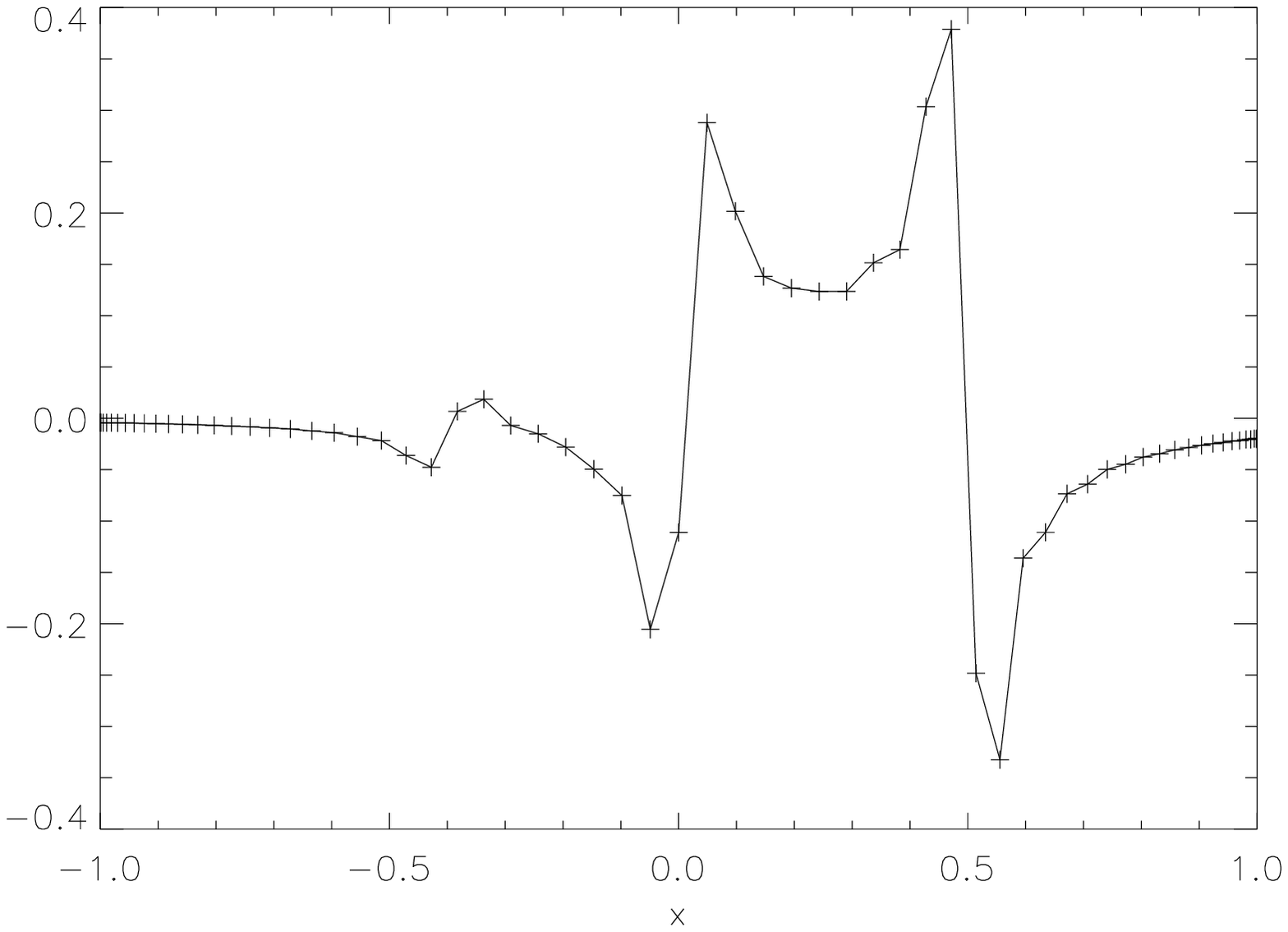}
      \hfill \vline \hfill
      \includegraphics[width=5cm]{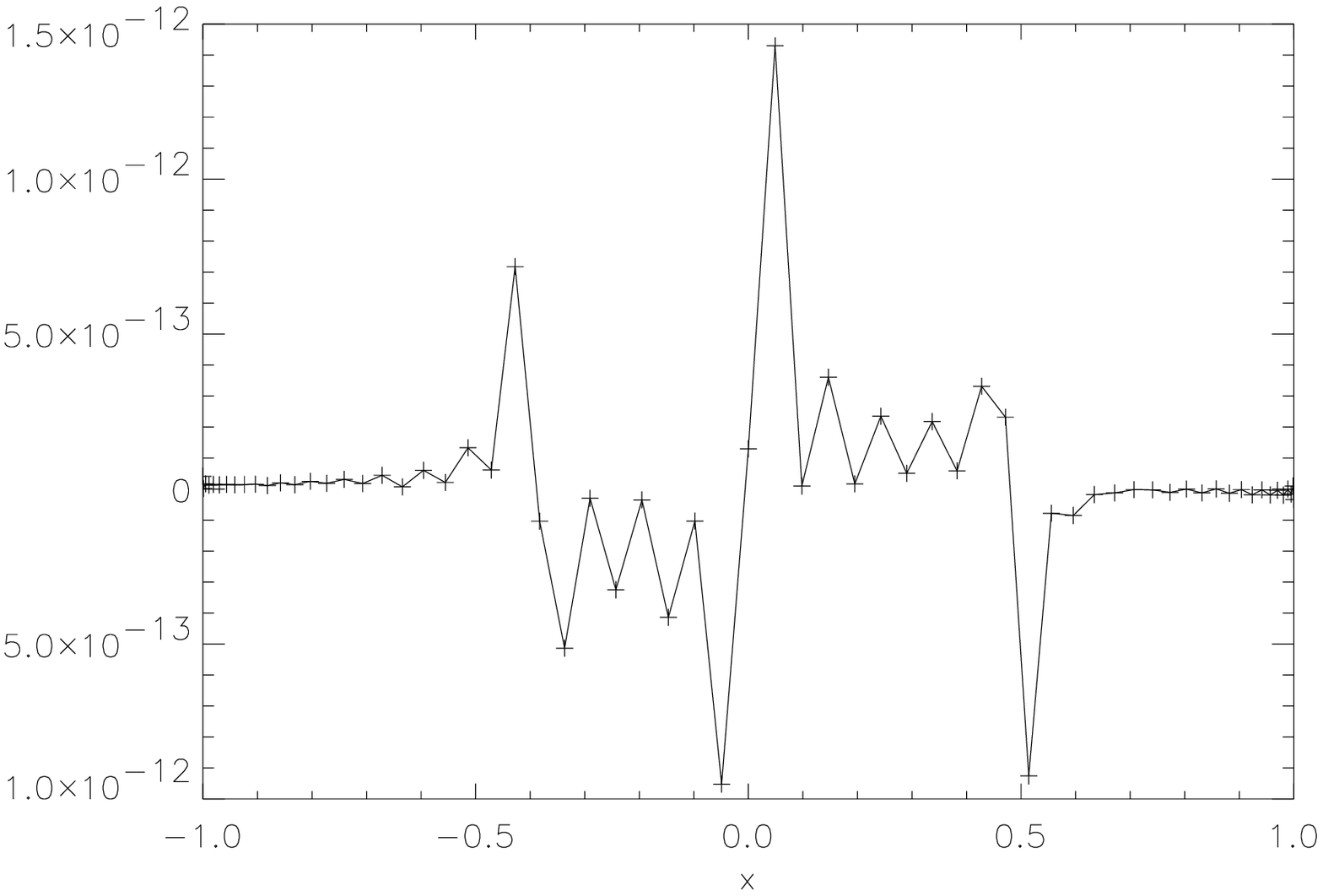}
      \hfill}
    \caption{\label{fig:divide} Results obtained with (right) and
      without (left) the regularizing projection}
  \end{center}
\end{figure}

Let $V_{M}$ be the space spanned by the polynomials with degree less
than the degree of $\omega$. Clearly, we have 
\begin{equation}
  \label{eq:jf_transverse}
  V_M \cap \omega[V_N] = \{0\}.  
\end{equation}
We need to change the projector $P=QQ^t=\hat Q_1\hat Q_1^t$ into a new
projector $\hat P = Q_1Q_1^t + Q_1SQ_2^t$ for some $(N+M)\times M$
matrix $S$ which has to be found from the fact that $\hat P$
annihilates $V_M$. This form of the new projector follows from the
fact that in the adapted basis given by the columns of $\hat Q$ a
general projector with image $\omega[V_N]$ has the matrix
representation
\[
\left(
  \begin{array}{c|c}
    \mathbf{1}_N&S\\
    \hline
    0&0
  \end{array}
\right).
\]
Let $U$ be a $(N+M)\times M$-matrix whose columns span $V_M$. Then $S$
has to satisfy the condition
\begin{equation}
  \label{eq:jf_condS}
  \left(\hat Q_1^t + S \hat Q_2^t\right) = 0.
\end{equation}
Note, that $Q_2^tU$ is a square matrix of dimension $M\times M$ which
is necessarily invertible. This is a consequence
of~\eqref{eq:jf_transverse}. Hence we get
\begin{equation}
  \label{eq:jf_S}
  S=\hat Q_1^t U \left(\hat Q_2^t U \right)^{-1}
\end{equation}
and consequently
\begin{equation}
  \label{eq:jf_newproj}
  \hat P = P\left( \mathbf{1}_{N+M} - U \left(\hat Q_2^t U \right)^{-1}
    \hat Q_2^t\right).
\end{equation}
Using this `improved' projection kills all the lower degree
polynomials and hence yields much smoother quotients. As an example
consider Fig.~\ref{fig:divide} where we have illustrated a
one-dimensional case with and without the use of the improved
projection.
The upper diagrams show the results obtained by dividing the function
$f = \omega(2x^3-7x^5+10\cos^2(5x))$ by
$\omega=(x-1/2)(x+2/5)(x-1/100)$. However, to simulate the real
problem we have contaminated $f$ with a low order polynomial (i.e. at
most quadratic in this case) with random coefficients of size at most
$\epsilon=10^{-2}$. The left diagram shows the quotient (indicated by
a line with crosses) without the use of the regularizing
projection. The line is the exact result. Clearly, the quotient is
influenced by the random deviation. The difference to the exact
quotient clearly shows the structure of $1/\omega$. The maximum error
is influenced by two sources: the size of $\epsilon$ i.e., the size of
the truncation error in the application we have in mind. The second
influence is the resolution, i.e. the number of grid points (or,
equivalently, the number of polynomials). With higher resolution
$1/\omega$ can be better resolved which results in higher spikes and
hence in bigger errors. On the right hand side the regularizing
projection has been used. Here, the deviation from the exact result is
not visible and, in fact, the error is at the level of
$10^{-12}$. Here, the error is not affected by $\epsilon$ because the
projection exactly kills any low order polynomial.
It appears that the regularizing projection redistributes the error in
a global manner so as to obtain a function which is `evenly' divisible
by $\omega$. 

As a final remark it should be noted that the way we have obtained the
regularizing projection does not in any way depend on the dimension of
the problem although we have essentially treated only the
one-dimensional case. In a similar way one can treat higher
dimensional cases by employing tensor bases and interpreting the
spaces of coefficients appropriately. The geometry behind the scenes
remains exactly the same.
\index{division~by~$\Omega$|)}
\index{pseudo-spectral~methods|)}

\section{Conclusion}

As we have seen the question of getting smooth initial data beyond the
conformal boundary of space-time is difficult. At the moment the only way
to obtain these data is via the Andersson--Chru\'sciel--Friedrich
procedure which has its limitations. Therefore, we should look for
other ways to get such initial data. There are two other possibilities for
this. One way is to understand the structure of the conformal
constraints and to find ways of formulating well-posed boundary value
problems for their direct solution. This is the route taken by
Butscher~\cite{butscher:_explor}.  

Another possible approach towards the construction of smooth initial
data is suggested by recent work by
J. Corvino~\cite{corvino00:_scalar_einst} (refined by Chru\'sciel and
Delay~\cite{chruscieldelay02:_exist}) who has constructed smooth
solutions to the constraint equations on an asymptotically Euclidean
space-like hypersurface which coincide exactly with Schwarzschild data
outside a compact set. The evolution of such data will contain at
least for some time an asymptotic region which is exactly
Schwarzschild. Therefore, it is possible, at least in principle, to
find hyperboloidal hypersurfaces in the time development of the data
on which hyperboloidal data are implied which are exactly
Schwarzschild outside a compact set and these data can trivially be
extended analytically beyond $\scri$. To obtain these data it is
necessary to either generalize Corvino's method to the hyperboloidal
setting and solve appropriate equations directly there or to generate
the hyperboloidal data numerically from Corvino's asymptotically flat
data. This requires an appropriate evolution scheme for Friedrich's
regular finite initial value problem at spatial
infinity~\cite{friedrich98:_gravit_fields} which so far is lacking.

Another issue which ultimately has to be thought about is the
inclusion of matter fields into the conformal framework. Including
matter fields is complicated because the Bianchi equation for the
rescaled Weyl tensor acquires a source term and this involves
derivatives of the stress-energy-tensor. So far there has been only
one case in which a system of matter fields coupled to the conformal
field equations has been studied
numerically~\cite{huebner94:_method_calc_sing_spti}.  While there is
no problem (at least in principle apart from complexity) to include
fundamental conformally invariant fields like the Maxwell, Yang-Mills
fields~\cite{friedrich91:_einst_maxwel_yang_mills} or other massless
fields like the conformally coupled scalar
field~\cite{huebner95:_gener_relat_scal_field} problems can arise for
massive fields like the Klein-Gordon field or ideal fluids. The reason
is that the matter equations are not conformally invariant so that a
regularization of the equations on $\scri$ by rescaling the fields
appropriately is unlikely. However, in the usual scenario of an
isolated system in which a source generates waves and sends them out
to infinity the matter region will not extend out to infinity so that
the singular behaviour on $\scri$ is not present. Even if one assumes
that matter fields extend to infinity one can still hope to be safe on
$\scri$ because in some cases like the massive Klein-Gordon field the
time evolution forces the field to decay exponentially towards $\scri$
so that the formal singularity is in fact not
there~\cite{winicour88:_massiv}.

To conclude it is probably safe to say that the conformal approach has
reached the status where we know that it is a reasonable and feasible
approach which offers some exciting possibilities but which also
provides us with problems (mostly of a numerical nature) which are
(apart from some idiosyncracies) very similar to those encountered in
other approaches. A joint effort of the numerical relativity community
seems to be called for in order to overcome these problems.

\index{initial~data|)}

\def\urlprefix{\relax}

\end{document}